\documentclass[prl,aps,showpacs,twocolumn]{revtex4}
\usepackage{epsf}
\usepackage[dvips]{color}
\newcommand {\be}{\begin{equation}}
\newcommand {\ee}{\end{equation}}
\newcommand {\bea}{\begin{eqnarray}}
\newcommand {\eea}{\end{eqnarray}}

\begin{document}

\title{Achieving a BCS transition in an atomic Fermi gas}
\author{L.~D. Carr$^1$$^*$, G.V.~Shlyapnikov$^{1,2,3}$$^{\dag}$ and Y. Castin$^1$}
\affiliation{\mbox{1. Laboratoire Kastler Brossel, Ecole Normale
Sup\'erieure, 24 rue Lhomond, 75231 Paris, France,}}
%\email[]{shlyap@goran.amolf.nl}
\affiliation{2. \mbox{FOM Institute for Atomic and Molecular
Physics, Kruislaan 407, 1098 SJ Amsterdam, The Netherlands}}
\affiliation{3. Russian Research Center, Kurchatov Institute,
Kurchatov Square, 123182 Moscow, Russia}

\date{\today}

\begin{abstract}
We consider a gas of cold fermionic atoms having two spin components
with interactions characterized by their $s$-wave scattering length
$a$. At positive scattering length the atoms form weakly bound bosonic
molecules which can be evaporatively cooled to undergo Bose-Einstein condensation,
whereas at negative scattering
length BCS pairing can take place.  It is
shown that, by adiabatically tuning the scattering length $a$ from
positive to negative values, one may transform the molecular
Bose-Einstein condensate into a highly degenerate atomic Fermi gas,
with the ratio of temperature to Fermi temperature $T/T_F \sim 10^{-2}$.
The corresponding critical final value of $k_{F}|a|$ which leads to the BCS
transition is found to be about one half, where $k_F$ is the Fermi momentum.
\end{abstract}


\maketitle

Much progress has been made in the achievement of increasingly
degenerate regimes of trapped atomic Fermi
gases~\cite{jin3,truscott1,schreck1,granade2002,roati2002,ohara1,hadzibabic2003}.
One of the major goals of studies of these systems is to observe a
transition to a paired-fermion superfluid state. In a recent MIT
experiment on a sympathetically cooled single component Fermi gas,
a system with $T/T_F=0.05$ and $3\times 10^7$ fermions was
realized~\cite{hadzibabic2003}, where $T_F$ is the Fermi
temperature. As trapped atomic gases present an essentially
impurity free system, with density, temperature, and interaction
strength all free parameters, they offer great opportunities for
investigation of fundamental theories of superfluid states.

An important new step in experiments has been the use of Feshbach
resonances, whereby one may tune the $s$-wave scattering length $a$
from positive to negative over many orders of
magnitude~\cite{vogels1}. This opens up the possibility of
investigating the Bose-Einstein condensate to
Bardeen-Cooper-Schrieffer (BEC-BCS)
crossover~\cite{nozieres1985,randeria1995,holland2001,ohashi2003}.
The BCS limit occurs for $k_F |a| \ll 1$, where $k_F$ is the Fermi
momentum and $a<0$.  In this case the fermion pair size is much
larger than the interatomic spacing.  The BEC limit occurs for a
positive scattering length much smaller than the interparticle
separation.  Then the fermions form weakly bound dimers of a size
$\sim a$, and the Bose-Einstein condensation of these composite
bosons may be described by the well-developed theory of trapped
BEC's~\cite{dalfovo1}. Recently, long-lived molecules have been 
created in a reversible, and therefore adiabatic
fashion from a degenerate Fermi gas by tuning the scattering
length~\cite{salomon2003}. This provides many
possibilities for evaporatively
cooling the molecules into the BEC regime. In contrast, a BCS transition
requires very low temperatures not yet obtained in two-component
Fermi gases.

In this Letter, we present a straightforward theoretical
analysis which shows that a deeply degenerate Fermi gas
may indeed be created from a molecular BEC. By adiabatically
changing the scattering length from positive to negative values,
the entropy is held constant, which leads to a strong decrease in temperature.
This suggests a novel method of cooling a Fermi gas to extremely low
temperatures. One can then reach a temperature
$T\sim 10^{-2}T_F$ and achieve a BCS transition.

Consider a harmonically trapped Fermi gas described by the grand
canonical ensemble with two equally populated spin states.
 The entropy of the gas above the critical temperature
is equal to the entropy of  an ideal
Fermi gas in a trap with mean field corrections. Omitting these corrections
the grand potential $\Omega$  is~\cite{diu1989}
\be \Omega=k_B T \int_0^{\infty} d\epsilon
\rho(\epsilon) \ln[1-n(\epsilon)]\, , \label{eqn:j}\ee with
$\rho(\epsilon)$ the density of states and \be n(\epsilon)=
1/\{\exp[\beta(\epsilon-\mu)]+1\}\ee the Fermi weighting
factor.  For a harmonic trap, the density of states for a
non-interacting two-component gas in the semi-classical limit
is \be \rho(\epsilon) =
\epsilon^2/(\hbar\omega)^3\equiv A \epsilon^2\,
.\label{eqn:dos}\ee Substituting Eq.~(\ref{eqn:dos}) into
Eq.~(\ref{eqn:j}) and integrating by parts, one obtains the
simplified integral $ \Omega=-(A/3)\int_0^{\infty} d\epsilon
\,\epsilon^3 n(\epsilon)$.  In the degenerate regime, given by
$\mu\gg k_B T$, the integral may be  expanded as \be
\Omega\simeq-\frac{A}{3}\left[\frac{\mu^4}{4}+\frac{\pi^2(k_B
T)^2\mu^2}{2} +\frac{7\pi^4 (k_B T)^4}{60}\right]
\label{eqn:j3}\, .\ee The entropy may then be obtained from the
relation~\cite{diu1989} \be S=-\partial_T \Omega
|_{\mu, \omega}\, ,\label{eqn:jtos}\ee giving \be S \simeq k_B
\frac{A}{3}\left[\pi^2 k_B T \mu^2 + \frac{7\pi^4 (k_B T)^3}{15}\right] \, .\ee
A relation between the total number of particles and the
chemical potential follows from the equation
$N=-\partial_{\mu}\Omega|_{T,\Omega}$ and reads
\be N \simeq (A/3)[\mu^3 +
\pi^2(k_B T)^2\mu] \, .\label{eqn:fmu}\ee 
So, to lowest order,  the chemical potential is $\mu=(3N/A)^{1/3}$
and the expression for the entropy becomes \be S=k_B
N \pi^2 T/T_F+{\cal O}(T^3)\, ,\label{eqn:sfermi}\ee  
where $T_F\equiv (3N)^{1/3}\,\hbar\omega$ is
the Fermi temperature of the non-interacting gas.  In the case of 
an interacting Fermi gas, the lowest order mean field correction 
in the Thomas-Fermi limit leads to the appearance of an extra 
multiple $(1+64k_F a/35\pi^2$) 
in the numerator of the right hand side of Eq.~(\ref{eqn:sfermi}).  
This gives a correction of less than 10 percent
to the entropy for $k_F|a|\leq 1/2$.  It may therefore be neglected in our calculations.  Equation (\ref{eqn:sfermi}) is valid
for $k_B T > \hbar\omega$ in an isotropic potential. Its validity
extends to much lower temperatures (although much larger than the
inverse density of states at the Fermi energy) for incommensurable
trap frequencies.

Consider the analogous calculation for a BEC of
weakly bound molecules at $a>0$.  We assume that the condensate is in the
weakly interacting regime, with $\eta \equiv n_{\mathrm{mol}}^{1/3}
a_{\mathrm{mol}} \ll 1$ where $n_{\mathrm{mol}}$ is the molecular density 
in the trap center and the
scattering length of two molecules $a_{\mathrm{mol}}$ is related to the
atomic scattering length $a$ by $a_{\mathrm{mol}} = 0.6
a$~\cite{petrov2003b}. Note that these weakly bound molecules exhibit a 
remarkable collisional stability at large values of 
$a$~\cite{salomon2003,petrov2003b}.  This stability 
originates from a strong decrease in the relaxation rate to deep 
bound states with increasing $a$, and should enable efficient 
evaporative cooling~\cite{petrov2003b}.  A small value of $\eta$ may be
reached in experiments by reducing the trap frequencies.

If $\eta \ll
1$, the mean field shift in the condensation temperature is
$\delta T_{\mathrm{BEC}}/T_{\mathrm{BEC}} = -3.25 \eta^{5/4}$ \cite{dalfovo1} and the
quantum depletion of the condensate is negligible so that
Bogoliubov theory may be used. It also implies \be k_B T_{\mathrm{BEC}}/E_B = 7.78 \eta^{5/2} \ll
1\label{eqn:cond1}\ee where $E_B\simeq \hbar^2/ma^2$ is the
molecular binding energy. This condition, together with $T<
T_{\mathrm{BEC}}$, ensures that the molecules may be treated as weakly
interacting bosons and any effect due to their thermal
dissociation is negligible.  The grand potential is \be \Omega=k_B T
\int_0^{\infty} d\epsilon \rho(\epsilon)
\ln\left(1-e^{-\beta\epsilon}\right)\, , \label{eqn:jb}\ee where
$\beta$ now refers to the temperature of the Bose gas.
The density of states for an interacting BEC in the
Bogoliubov approximation may be calculated directly from the
Bogoliubov Hamiltonian as
$\rho(\epsilon)=\mathrm{Tr}[\delta(\epsilon-\hat{H}_{bog})]$.  In
the Thomas-Fermi limit, which holds for a molecular condensate 
with chemical potential $\mu_{\mathrm{mol}} \gg \hbar \omega$,
the density of states may be approximated semi-classically by \be
\rho(\epsilon)\simeq \int\int
\frac{d\vec{r}d\vec{p}}{(2\pi\hbar)^3}
\:\delta[\epsilon-\epsilon_{bog}(\vec{p},\vec{r})]\,
,\label{eqn:dos2}\ee  where the energy of Boguliubov excitations is (see~\cite{dalfovo1} and references therein) \be \epsilon_{bog} \equiv
\left\{\sqrt{p^2/2m_{\mathrm{mol}}\left[p^2/2m_{\mathrm{mol}}+2g_{\mathrm{mol}}n_0(r)\right]}
\quad|r|\leq R \atop{p^2/2m_{\mathrm{mol}}+\frac{1}{2}m_{\mathrm{mol}}\omega^2r^2
-\mu_{\mathrm{mol}} \quad\,\,\,\,\,\,\,\,\,\, |r|>R\, .}\right .\, \ee
Here $m_{\mathrm{mol}}=2m$ is the
mass of a molecule and
\be
n_0(r)=(\mu_{\mathrm{mol}}/g_{\mathrm{mol}})(1-r^2/R^2)\,  
\label{eqn:tf}\ee is the
Thomas-Fermi density profile, with $R\equiv(2\mu_{\mathrm{mol}}/m_{\mathrm{mol}}\omega^2)^{1/2}$.   
The coupling constant
for the molecule--molecule interaction is
$g_{\mathrm{mol}}=4\pi\hbar^2 a_{\mathrm{mol}}/m_{\mathrm{mol}}$.
The exact calculation of Eq.~(\ref{eqn:dos2}) yields \bea
\rho(\epsilon)=[\mu_{\mathrm{mol}}^2/(\pi\hbar^3\omega^3)]
\{2\sqrt{2}z\tanh^{-1}[\sqrt{2z}/(1+z)
]\nonumber\\
+4z^{3/2}-\sqrt{2}z^2 [\pi+2\tan^{-1}
((1-z)/\sqrt{2z})]\nonumber\\
+(1+z)^2 [\theta_0 -\sin(4\theta_0)/4]\}\, ,\,\,\,\,\eea where
$z\equiv \epsilon/\mu_{\mathrm{mol}}$ is the rescaled energy and
$\theta_0\equiv\cos^{-1}(1/\sqrt{1+z})$.  In the limit $k_B
T\gtrsim \mu_{\mathrm{mol}}$, \be
\rho(\epsilon)=\frac{\mu_{\mathrm{mol}}^2}{(\hbar\omega)^3}\left[\frac{
z^2}{2} +z+\mathcal{O}(z^{0})\right]\, .\label{eqn:lim2}\ee Given
the density of states, the entropy can be calculated from Eq.~(\ref{eqn:jtos}) 
and~(\ref{eqn:jb}).  The resulting expression is \be S=k_B
\frac{N_{\mathrm{mol}}}{\zeta(3)}\left(\frac{T}{T_{\mathrm{BEC}}}\right)^3
G(\beta\mu_{\mathrm{mol}})\, ,\label{eqn:sbose}\ee where
$\zeta$ is the Riemann zeta function, with $\zeta(3)=1.202\ldots$,
$N_{\mathrm{mol}}=N/2$ is the number of molecules and \be G(u)
\equiv u^3 \int_0^{\infty}dz f(z)\left[\frac{u
z}{e^{u z}-1}- \ln(1-e^{-u z})\right] \,
,\label{eqn:G}\ee with $u\equiv\beta\mu_{\mathrm{mol}}$, and $f(z)\equiv
[(\hbar\omega)^3/\mu_{\mathrm{mol}}^2]\rho(z\mu_{\mathrm{mol}})$ so as to make the
units explicit. Equation~(\ref{eqn:G}) may be integrated
numerically, or explicitly in the limit of Eq.~(\ref{eqn:lim2}),
in which case \be S=k_B N_{\mathrm{mol}}\left(
\frac{T}{T_{\mathrm{BEC}}}\right)^3\left(\frac{2 \pi^4 }{45 \zeta (3)} +
3\frac{\mu_{\mathrm{mol}}}{k_B T}\right)\, \label{eqn:expand}\ee This
expansion is accurate to within 10\% for $k_B T/\mu_{\mathrm{mol}} \geq 1/10$. In
the above expression for the entropy, the condensation temperature
for a non-interacting Bose gas, $k_B T_{\mathrm{BEC}}=\hbar\omega
[N_{\mathrm{mol}}/\zeta(3)]^{1/3}$ was used. Setting the entropy of the
molecular condensate and of the Fermi gas equal one obtains the
final temperature of the Fermi gas: \be
\left(\frac{T}{T_F}\right)_{\mathrm{final}}=\frac{G(\beta\mu_{\mathrm{mol}})}{2\pi^2\zeta(3)}
\left(\frac{T}{T_{\mathrm{BEC}}}\right)_{\mathrm{initial}}^3
\label{eq:final_temperature} \ee
 where $\beta$ refers to the initial temperature.

%%%%%%%%%%% figure 1 %%%%%%%%%%%
%
\begin{figure}[t]
\begin{center}
\epsfxsize=8cm  \leavevmode \epsfbox{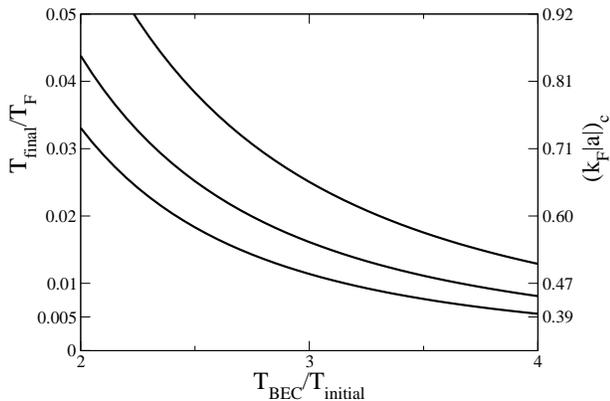}
\caption{\label{fig:1}  
Final temperature of the Fermi gas (left axis) 
and critical condition for a BCS transition  (right
axis) as a function of  the initial temperature of the Bose
condensed molecular gas that has been transformed to  a normal
attractive Fermi gas via adiabatic switching  of the scattering
length. The curves, from upper to lower, correspond to
$\mu_{\mathrm{mol}}/k_B T_{\mathrm{BEC}}=$ $1$, $1/2$,
and $1/4$, where $\mu_{\mathrm{mol}}$ is the molecular chemical
potential.}
\end{center}
\end{figure}

 In the strongly degenerate regime, evaporative cooling of a Fermi gas becomes
difficult, so that until recently a maximum degeneracy of $T/T_F \sim 0.2$ appeared to be a lower
limit, preventing the observation of the BCS transition.
 The adiabatic tuning of a molecular condensate suggests an
alternate method to cooling the fermions directly.  According to
Eqs.~(\ref{eqn:sfermi}) and~(\ref{eqn:expand}), the entropy is
proportional to $T$ for fermions and $T^3$ for bosons.  The limits
of evaporative cooling for a  Thomas-Fermi Bose condensed gas are given by
$\mu\simeq k_B T$.  Thus the final degeneracy that may be obtained
by adiabatic switching of an evaporatively cooled molecular BEC is
given by Eq.~(\ref{eq:final_temperature}),
with $G(\beta\mu_{\mathrm{mol}})=G(1)\simeq 8.32$.
Larger values of  $\beta\mu_{\mathrm{mol}}$ are in principle achievable, {\it e.g.}
by sympathetic cooling. In Fig.~\ref{fig:1} is shown the final temperature as
a function of the initial temperature
for various interaction strengths given by the
ratio $\mu_{\mathrm{mol}}/k_B T_{\mathrm{BEC}}$.

The lowest experimentally achieved temperature of a degenerate
Fermi gas to date is $T/T_F=0.05$~\cite{hadzibabic2003}.  This
temperature may be obtained by adiabatic switching of a molecular
BEC of temperature $T=0.5\, T_{\mathrm{BEC}}$
with $\mu_{\mathrm{mol}}\simeq k_B T$,
which is routinely achieved for atomic BEC's. A temperature of
$T=0.25 \,T_{\mathrm{BEC}}$ with  $\mu_{\mathrm{mol}}\simeq k_B T$, 
as for example achieved
for  an atomic Bose condensed gas in
Ref.~\cite{chevy2002}, would give $T/T_F=5\times 10^{-3}$.

This offers exciting possibilities for obtaining deeply degenerate
Fermi regimes, such as observing the BCS transition in the weakly
interacting regime $k_F|a|\ll 1$.  The  corresponding critical condition is
obtained by substituting into Eq.~(\ref{eq:final_temperature})
the BCS critical temperature calculated in Ref.~\cite{gorkov1}:
\be T_{\mathrm{BCS}}/T_F = \alpha \exp(-\pi/2k_F|a|)\,
,\label{eqn:bcsTc}\ee
where  $\alpha\equiv e^\gamma\pi^{-1}(2/e)^{7/3}\simeq 0.277$,
$\gamma$ being Euler's constant.
Strictly speaking, in an isotropic harmonic potential,
Eq.~(\ref{eqn:bcsTc}) is only valid in the limit $k_B T_{\mathrm{BCS}}\gg
\hbar \omega$. However, for $k_F|a|=1/2$ or $1/3$, it gives a
result accurate to 30\%
for $k_B T_{\mathrm{BCS}}\simeq 2\hbar \omega$~\cite{baranov3}. In
the right axis of Fig.~\ref{fig:1} is shown the critical value $(k_F|a|)_c$
ensuring $T_{\rm final}=T_{\mathrm{BCS}}$, as a function
of $T_{\mathrm{BEC}}/T$:
$(k_F|a|)_c$ is on the order of one half, a regime
already accessible by present experiments.
It should be noted that our treatment is exactly valid only in the
regime $k_F |a| \ll 1$, as we have used BCS theory. The values of
$(k_F |a|)_c$ that we find are not extremely small compared to
unity; we therefore expect corrections to both the equation of
state of the normal phase and the critical temperature. Our result
must therefore be considered as an estimate.

Fermi gases have been predicted to be sensitive to heating due to
particle losses, much more so than Bose
condensates~\cite{timmermans2}. One may ask what the limit is on the
accessible final temperature set by this heating during the
adiabatic tuning of $a$. Assume a loss rate
$\gamma_{\mathrm{loss}}$ of atoms due mainly to collisions with
the background gas, and let $t$ be the duration of the adiabatic
ramp. Using the derivation of Ref.~\cite{idziaszek2002}, and
assuming that $\gamma_{\mathrm{loss}} t \ll 1$ and
$(T/T_F)^2 \ll 1$, the temperature increase due to loss may be
estimated as $(T/T_F)_{\mathrm{increase}}\sim\sqrt{\gamma_{\mathrm{loss}} t/(2
\pi^2)}$. One may take  $t\sim 1/[\gamma_{\mathrm{coll}}(T/T_F)^2]$, where
$\gamma_{\mathrm{coll}}=n\sigma v_F$ is the classical
collision rate  involving the scattering cross-section
$\sigma\simeq 4\pi a^2$.  Here $n$ is the total density, 
$v_F\equiv \hbar k_F / m$ is the Fermi velocity,
and a factor of $(T/T_F)^2$ is included to account for
the Pauli blocking for each of the two components~\cite{ohara1}.
Then, supposing $(T/T_F)_{\mathrm{final}}\sim 10^{-2}$,
one finds the condition \be
\left[\gamma_{\mathrm{loss}}/(2\pi^2\gamma_{\mathrm{coll}})\right]^{1/4}\lesssim
(T/T_F)_{\mathrm{final}}\sim 10^{-2} \label{eqn:cri}\ee in order to avoid a
loss-induced increase in temperature during the switching time.
For the experiment of Ref.~\cite{salomon2003}, $\gamma_{\mathrm{loss}} < 10^{-2}$
s${}^{-1}$, and Eq.~(\ref{eqn:cri}) is satisfied for $a\gtrsim
1000 a_0$ and  $n\gtrsim 5\times 10^{12}$ cm$^{-3}$, where $a_0$ is the Bohr radius.

In order to satisfy the criterion of adiabaticity,
thermal equilibrium must be maintained while tuning the scattering length from positive
to negative values.
This requires that the thermalization rate, {\it i.e.}, the rate of elastic collisions,
be much larger than the rate of change of $a$ \cite{proviso}.
For $a>0$, thermalization is due to collisions between non-condensed molecules.
As the size of the non-condensed cloud is close to the size of the condensate
 for $k_B T \sim \mu_{\mathrm{mol}}$,
the effective collision rate is
thus
$\gamma_{\mathrm{eff}}=\gamma_{\mathrm{coll}}^{\mathrm{mol}}\, N_T/N_{\mathrm{mol}}$,
where $N_T/N_{\mathrm{mol}}$ is the non-condensed fraction and
$\gamma_{\mathrm{coll}}^{\mathrm{mol}}= n_{\mathrm{mol}} \sigma_{\mathrm{mol}} v_T$,
with $v_T$ the thermal velocity
and $\sigma_{\mathrm{mol}}=8\pi a_{\mathrm{mol}}^2$ the scattering
cross section between two molecules.  For $T/T_{\mathrm{BEC}}=1/2$, the non-condensed fraction
is $(T/T_{\mathrm{BEC}})^3=1/8$~\cite{dalfovo1}.
Assuming $a\sim 1000 a_0$ and the mass of $^6$Li, even for
molecular densities $n_{\mathrm{mol}}$ as small
as $10^{12}$ cm$^{-3}$, this gives a minimum thermalization time of  $\sim 30$ ms.
For $a<0$, the rate of thermalization is strongly reduced by Pauli blocking, and
one has $\gamma_{\mathrm{eff}}=\gamma_{\mathrm{coll}}(T/T_F)^2$~\cite{ohara1}.
At the critical value of $k_F|a|\sim 1/2$,  and for a total density
of  $5\times 10^{12}$ cm$^{-3}$,  adiabaticity requires
that the change of the scattering length occurs on a time $t>300$ ms.
 This time is longer than the inverse oscillation frequency
of lithium atoms in typical magnetic traps, so that the change in $a$ will not
induce macroscopic oscillations in the gas.

It is thus far not possible to obtain a perfect balance of spin
states in a Fermi degenerate gas~\cite{salomon2003}. It may
therefore be supposed that a small fraction of fermions will
remain unpaired for positive scattering lengths, due to a lack of
partners, and will coexist with the molecular condensate.  
%As the
%fermions are in a single spin state $s$-wave approximation they 
%interact only with the molecules.  The  mean field plus trapping potential
%experienced by an unpaired fermion is
%$m\omega^2 r^2/2 + g_{am}\rho_{\mathrm{mol}}(r)$, where
%the atom-molecule coupling constant is $g_{am}=0.9
%g=0.9\times 4\pi\hbar^2 a/m$~\cite{petrov2003b} and the molecular mean field, in the
%Thomas-Fermi limit, is given by
%Eq.~(\ref{eqn:tf}), with
%$g_{\mathrm{mol}}=0.3 g$.  They therefore experience an effective potential
%$U_{\mathrm{eff}}(r)=-2.5\,m\omega^2r^2 +3\mu_{\mathrm{mol}}$. This causes a spatial
%separation between atoms and molecules, with atoms pushed to the
%outside of the molecular mean field. 
Fermions have an entropy
proportional to $T$, as opposed to $T^3$ in the case of bosons. So
even a small number of atoms could be expected to dominate the
total entropy of the atom-molecule mixture at low temperatures,
thereby interfering with the proposed cooling scheme.
This problem can
be avoided by simply removing the excess fermions from the system
for $a>0$ with a correctly tuned laser pulse, which requires that
their excitation frequencies are sufficiently different from those
of the molecules~\cite{jinPrivateCommunication2003}.

In conclusion, we have shown that adiabatic switching
of a molecular BEC allows one to obtain a deeply
degenerate Fermi gas of temperatures on the order of $T\sim
10^{-2} T_F$.  This
suggests a way to achieve a BCS transition in the weakly
interacting regime.  It is important to
note that our thermodynamic approach does not require a knowledge
of the  equation of state of the system in the intermediate strong coupling
regime where $k_F|a| \gtrsim 1$.

We thank Christophe Salomon, Deborah Jin and Carlos Lobo for useful comments.
This work was financially supported by NSF grant no.~MPS-DRF
0104447, the French Minist\`ere des Affaires Etrang\`eres, by the
Dutch Foundations NWO and FOM, by INTAS, and by the Russian
Foundation for Fundamental Research.  Laboratoire Kastler Brossel
is a research unit of Universit\'e Pierre et Marie Curie and Ecole
Normale Sup\'erieure, associated with CNRS (UMR 8552).

{\it Note added}: Since submission of this paper, long-lived BEC's of 
weakly bound molecules have been observed~\cite{Greiner03}, and a strongly interacting fermionic condensate in the BEC-BCS crossover regime
was created~\cite{regal2004}.


\vspace*{-6mm}

\begin{thebibliography}{10}

\bibitem[*]{byline1}
Present address: JILA, National Institute of Standards 
and Technology and Physics Department, University of Colorado,
Boulder, CO 80309-0440

\bibitem[$\dag$]{byline2}
Present address: Laboratoire de Physique Th\'eorique et Mod\`eles Statistiques,
Universit\'e Paris Sud, B\^at. 100, 91405 Orsay Cedex, France;
and Van der Waals - Zeeman Institute, University of Amsterdam,
Valckenierstraat 65/67, 1018 XE Amsterdam, The Netherlands 

\bibitem{jin3}
B. Demarco and D.~S. Jin, Science {\bf 285},  1703  (1999).

\bibitem{truscott1}
A. G. Truscott, K. E. Strecker, W. I. McAlexander, G. Partridge, and R. G. Hulet, Science {\bf 291},  2570  (2001).

\bibitem{schreck1}
F. Schreck, L. Khaykovich, K. L. Corwin, G. Ferrari, T. Bourdel, J. Cubizolles, and C. Salomon, Phys. Rev. Lett. {\bf 87},  080403 (2001).

\bibitem{granade2002}
S.~R. Granade, M.~E. Gehm, K.~M. O'Hara, and J.~E. Thomas, 
Phys. Rev. Lett.
  {\bf 88},  120405  (2002).

\bibitem{roati2002}
G. Roati, F. Riboli, G. Modugno, and M. Inguscio, Phys. Rev. Lett.
{\bf 89},
  150403  (2002).

\bibitem{ohara1}
K. M. O'Hara, S. L. Hemmer, M. E. Gehm, S. R. Granade, and J. E. Thomas, Science {\bf 298},  2179
(2002).

\bibitem{hadzibabic2003}
Z. Hadzibabic, S. Gupta, C. A. Stan, C. H. Schunck, M. W. Zwierlein, K. Dieckmann, and W. Ketterle, Phys. Rev. Lett. {\bf 91}, 160401 (2003).

\bibitem{vogels1}
J. M. Vogels, C. C. Tsai, R. S. Freeland, S. J. J. M. F. Kokkelmans, B. J. Verhaar, and D. J. Heinzen, Phys. Rev. A {\bf 56},  R1067  (1997).

\bibitem{nozieres1985}
P. Nozi\`eres and S. Schmitt-Rink, J. Low Temp. Phys. {\bf 59},
195  (1985).

\bibitem{randeria1995}
M. Randeria, {\em Bose-Einstein Condensation} (Cambridge
University Press,
  U.K., 1995), Chap.~15, pp.\ 355--392.

\bibitem{holland2001}
M. Holland, S.~J. J. M.~F. Kokkelmans, M.~L. Chiofalo, and R. Walser, 
Phys. Rev. Lett. {\bf 87},  120406  (2001).

\bibitem{ohashi2003}
Y. Ohashi and A. Griffin, Phys. Rev. A {\bf 67},  033603  (2003).

\bibitem{proviso}
One may wonder if the tuning time of $a$ should not also be longer than, e.g., the
formation time of the BCS state, which could exceed the collision time.
However, in our scheme, the gas is initially a Bose condensate
of molecules, so that no formation of a superfluid component from scratch
is required.

\bibitem{dalfovo1}
F. Dalfovo, S. Giorgini, L.~P. Pitaevskii, and S. Stringari, Rev.
Mod. Phys. {\bf 71},  463  (1999).

\bibitem{salomon2003}
J. Cubizolles, T. Bourdel, S. J. J. M. F. Kokkelmans, G. V. Shlyapnikov, and C. Salomon, Phys. Rev. Lett. {\bf 91}, 240401 (2003); S. Jochim, M. Bartenstein, A. Altmeyer, G. Hendl, C. Chin, J. H. Denschlag, and R. Grimm, Phys. Rev. Lett. {\bf 91}, 240402 (2003); K.E. Strecker, G.B. Partridge, R.G. Hulet, Phys. Rev. Lett. {\bf
91}, 080406 (2003); C.A. Regal, M. Greiner, and D.S. Jin, e-print
cond-mat/0308606 (2003).

\bibitem{urban2003}
M. Urban and P. Schuck, Phys. Rev. A {\bf 67},  033611  (2003).

\bibitem{diu1989}
B. Diu, C. Guthman, D. Lederer, and B. Roulet, {\em Physique
Statistique}
  (Hermann, Paris, France, 1989).

\bibitem{petrov2003b}
D.~S. Petrov, C. Salomon, and G.V. Shlyapnikov, e-print cond-mat/0309010 (2003).

\bibitem{gorkov1}
L.~P. Gor'kov and T.~K. Melik-Barkhudarov, Sov. Phys. JETP {\bf
13},  1018
  (1961).

\bibitem{baranov3}
M.~A. Baranov and D.~S. Petrov, Phys. Rev. A {\bf 58},  R801
(1998).

\bibitem{chevy2002}
F. Chevy {\it et~al.}, Phys. Rev. Lett. {\bf 88},  250402  (1997).

\bibitem{jinPrivateCommunication2003}
Deborah Jin, JILA, Univ. of Colorado, private communication
(2003).

\bibitem{timmermans2}
E. Timmermans, Phys. Rev. Lett. {\bf 87},  240403  (2001); 
L.~D. Carr, T. Bourdel, and Y. Castin, Phys. Rev. A in press (2004).

\bibitem{idziaszek2002}
Z. Idziaszek, L. Santos, M. Baranov, and M. Lewenstein, 
Phys. Rev. A {\bf 67}, 041403  (2003).

\bibitem{Greiner03}
M. Greiner, C. Regal, and D.~S. Jin, Nature {\bf 426}, 537 (2003);
S. Jochim, M. Bartenstein, A. Altmeyer, S. Riedl, C. Chin, J. H. Denschlag, and R. Grimm, Science {\bf 302}, 2101 (2003); 
M. W. Zwierlein, C. A. Stan, C. H. Schunck, S. M. F. Raupach, S. Gupta, Z. Hadzibabic, and W. Ketterle, Phys. Rev. Lett. {\bf 91}, 250401 (2003); 
C. Salomon, private communication (2003); 
M. Bartenstein, A. Altmeyer, S. Riedl, S. Jochim, C. Chin, J. H. Denschlag, R. and Grimm, e-print cond-mat/0401109 (2004).

\bibitem{regal2004}
C.~A. Regal, M. Greiner, and D.~S. Jin, Phys. Rev. Lett. {\bf 92} 040403 (2004).

\end{thebibliography}
\end{document}